\documentclass[11pt]{article}
\usepackage[margin=1in]{geometry}
\usepackage[numbers,sort&compress]{natbib}
\usepackage{ff-manuscript}
\usepackage[hidelinks]{hyperref}
\newcommand{\papertitle}{Pretrained Tabular Features and the
Measurement of Financial Proximity}
\newcommand{\paperauthors}{Marcus Gawronsky} 
\newcommand{\paperdate}{23 November 2026}

  \ffDeclareValue{quarterly-fernandez-joint-effective-neighbours}{836.811}
\ffDeclareNumber{quarterly-fernandez-joint-effective-neighbours}{836.811295002107}
\ffDeclareValue{quarterly-fernandez-joint-premium}{0.003}
\ffDeclareNumber{quarterly-fernandez-joint-premium}{0.0029778958666073128}
\ffDeclareValue{quarterly-fernandez-joint-rho}{0.860}
\ffDeclareNumber{quarterly-fernandez-joint-rho}{0.8599956224156718}
\ffDeclareValue{quarterly-fernandez-joint-rho-bootstrap-draws}{999}
\ffDeclareNumber{quarterly-fernandez-joint-rho-bootstrap-draws}{999}
\ffDeclareValue{quarterly-fernandez-joint-rho-bootstrap-high}{0.964}
\ffDeclareNumber{quarterly-fernandez-joint-rho-bootstrap-high}{0.9644802078832753}
\ffDeclareValue{quarterly-fernandez-joint-rho-bootstrap-low}{0.225}
\ffDeclareNumber{quarterly-fernandez-joint-rho-bootstrap-low}{0.22465753140201805}
\ffDeclareValue{quarterly-panel-issuer-quarters}{54,906}
\ffDeclareNumber{quarterly-panel-issuer-quarters}{54906}
\ffDeclareValue{quarterly-panel-quarters}{60}
\ffDeclareNumber{quarterly-panel-quarters}{60}
\ffDeclareValue{quarterly-primary-issuers}{878}
\ffDeclareNumber{quarterly-primary-issuers}{878}
\ffDeclareValue{quarterly-primary-population-issuers}{990}
\ffDeclareNumber{quarterly-primary-population-issuers}{990}
\ffDeclareValue{quarterly-primary-roster}{967}
\ffDeclareNumber{quarterly-primary-roster}{967}
\ffDeclareValue{apt-apt-draws}{9,999}
\ffDeclareNumber{apt-apt-draws}{9999}
\ffDeclareValue{apt-apt-exiting}{123}
\ffDeclareNumber{apt-apt-exiting}{123}
\ffDeclareValue{apt-apt-gmv}{0.133}
\ffDeclareNumber{apt-apt-gmv}{0.13292555720828203}
\ffDeclareValue{apt-apt-issuers}{684}
\ffDeclareNumber{apt-apt-issuers}{684}
\ffDeclareValue{apt-apt-mantel-draws}{999}
\ffDeclareNumber{apt-apt-mantel-draws}{999}
\ffDeclareValue{apt-apt-roster}{967}
\ffDeclareNumber{apt-apt-roster}{967}
\ffDeclareValue{apt-apt-test_issuer_months}{36,061}
\ffDeclareNumber{apt-apt-test_issuer_months}{36061}
\ffDeclareValue{apt-apt-test_months}{60}
\ffDeclareNumber{apt-apt-test_months}{60}
\ffDeclareValue{apt-apt-train_months}{60}
\ffDeclareNumber{apt-apt-train_months}{60}
\ffDeclareValue{apt-cal-over-raw-better-holm}{1}
\ffDeclareNumber{apt-cal-over-raw-better-holm}{1}
\ffDeclareValue{apt-cal-over-raw-count}{10}
\ffDeclareNumber{apt-cal-over-raw-count}{10}
\ffDeclareValue{apt-cal-over-raw-positive}{5}
\ffDeclareNumber{apt-cal-over-raw-positive}{5}
\ffDeclareValue{apt-capm-ci}{[-27.6, 9.0]}
\ffDeclareValue{apt-capm-gain}{-8.6}
\ffDeclareNumber{apt-capm-gain}{-8.591470754555479}
\ffDeclareValue{apt-certified-max}{0.000}
\ffDeclareNumber{apt-certified-max}{0.0}
\ffDeclareValue{apt-characteristics-ci}{[-2.7, 0.1]}
\ffDeclareValue{apt-characteristics-gain}{-1.3}
\ffDeclareNumber{apt-characteristics-gain}{-1.2670366085839961}
\ffDeclareValue{apt-characteristics-lambda}{0.17}
\ffDeclareNumber{apt-characteristics-lambda}{0.1686367182433359}
\ffDeclareValue{apt-comparisons}{74}
\ffDeclareNumber{apt-comparisons}{74}
\ffDeclareValue{apt-direct-rsq-hist}{0.12}
\ffDeclareNumber{apt-direct-rsq-hist}{0.12023438450418944}
\ffDeclareValue{apt-direct-rsq-xbrl}{0.25}
\ffDeclareNumber{apt-direct-rsq-xbrl}{0.2521284670890368}
\ffDeclareValue{apt-error-margin-ratio}{250}
\ffDeclareNumber{apt-error-margin-ratio}{250}
\ffDeclareValue{apt-exp-eff-max}{837}
\ffDeclareNumber{apt-exp-eff-max}{836.811295002107}
\ffDeclareValue{apt-exp-eff-min}{665}
\ffDeclareNumber{apt-exp-eff-min}{665.3750287629712}
\ffDeclareValue{apt-exp-uniform-max}{0.20}
\ffDeclareNumber{apt-exp-uniform-max}{0.20386611565973647}
\ffDeclareValue{apt-exp-uniform-min}{0.08}
\ffDeclareNumber{apt-exp-uniform-min}{0.07893515479243496}
\ffDeclareValue{apt-exposure-ci}{[-8.2, 6.2]}
\ffDeclareValue{apt-exposure-gain}{-1.0}
\ffDeclareNumber{apt-exposure-gain}{-1.0274027306002853}
\ffDeclareValue{apt-exposure-lambda}{0.51}
\ffDeclareNumber{apt-exposure-lambda}{0.5073879206183373}
\ffDeclareValue{apt-fm-better-holm}{0}
\ffDeclareNumber{apt-fm-better-holm}{0}
\ffDeclareValue{apt-fm-better-point}{2}
\ffDeclareNumber{apt-fm-better-point}{2}
\ffDeclareValue{apt-fm-mantel-cal-mean}{0.20}
\ffDeclareNumber{apt-fm-mantel-cal-mean}{0.1955599507960449}
\ffDeclareValue{apt-fm-mantel-raw-mean}{0.11}
\ffDeclareNumber{apt-fm-mantel-raw-mean}{0.10861940386365682}
\ffDeclareValue{apt-fm-rsq-max}{0.22}
\ffDeclareNumber{apt-fm-rsq-max}{0.22034293600802457}
\ffDeclareValue{apt-fm-rsq-min}{0.10}
\ffDeclareNumber{apt-fm-rsq-min}{0.09600515478400073}
\ffDeclareValue{apt-fm-spatial-count}{16}
\ffDeclareNumber{apt-fm-spatial-count}{16}
\ffDeclareValue{apt-fm-worse-holm}{3}
\ffDeclareNumber{apt-fm-worse-holm}{3}
\ffDeclareValue{apt-generated-draws}{999}
\ffDeclareNumber{apt-generated-draws}{999}
\ffDeclareValue{apt-industry-ci}{[26.2, 40.5]}
\ffDeclareValue{apt-industry-gain}{33.3}
\ffDeclareNumber{apt-industry-gain}{33.34646296165727}
\ffDeclareValue{apt-industry-gmv}{0.131}
\ffDeclareNumber{apt-industry-gmv}{0.13135277611030358}
\ffDeclareValue{apt-industry-lambda}{0.45}
\ffDeclareNumber{apt-industry-lambda}{0.44624429409873795}
\ffDeclareValue{apt-intercept-boundary}{15}
\ffDeclareNumber{apt-intercept-boundary}{15}
\ffDeclareValue{apt-intercept-fits}{15}
\ffDeclareNumber{apt-intercept-fits}{15}
\ffDeclareValue{apt-issuers}{878}
\ffDeclareNumber{apt-issuers}{878}
\ffDeclareValue{apt-jaccard-max}{0.036}
\ffDeclareNumber{apt-jaccard-max}{0.035741037988229}
\ffDeclareValue{apt-joint-mantel-exposure}{0.06}
\ffDeclareNumber{apt-joint-mantel-exposure}{0.0608995457611273}
\ffDeclareValue{apt-joint-premium}{0.003}
\ffDeclareNumber{apt-joint-premium}{0.0029778958666073128}
\ffDeclareValue{apt-joint-rho}{0.860}
\ffDeclareNumber{apt-joint-rho}{0.8599956224156718}
\ffDeclareValue{apt-joint-rho-ci}{[0.225, 0.964]}
\ffDeclareValue{apt-margin-median-max}{1.4}
\ffDeclareNumber{apt-margin-median-max}{1.3771339273745}
\ffDeclareValue{apt-ols-premium-with}{0.002}
\ffDeclareNumber{apt-ols-premium-with}{0.0022124218207906594}
\ffDeclareValue{apt-ols-premium-with-ci}{[-0.001, 0.005]}
\ffDeclareValue{apt-ols-premium-without}{0.016}
\ffDeclareNumber{apt-ols-premium-without}{0.016404622979702105}
\ffDeclareValue{apt-ols-premium-without-ci}{[0.014, 0.018]}
\ffDeclareValue{apt-rms-error-min}{0.034}
\ffDeclareNumber{apt-rms-error-min}{0.034079494170938}
\ffDeclareValue{apt-roster}{967}
\ffDeclareNumber{apt-roster}{967}
  \ffDeclareValue{matched-observations}{51,782}
\ffDeclareValue{matched-broad-gain}{0.0085}
\ffDeclareValue{matched-factor-gain}{0.0013}
\ffDeclareValue{matched-factor-interval}{[0.0003, 0.0024]}
\ffDeclareValue{matched-historical-beta-r2}{0.25}
\ffDeclareValue{matched-historical-beta-n}{878}
\ffDeclareValue{matched-historical-value-r2}{93}
\ffDeclareValue{matched-historical-value-n}{960}
\ffDeclareValue{matched-inertia}{46.15}
\ffDeclareValue{matched-valuation}{0.93}
\ffDeclareValue{matched-overlap}{0.053}
\ffDeclareValue{matched-projected-gain}{-0.0116}

\title{\papertitle}
\author{\paperauthors}
\date{\paperdate}
\begin{document}
\maketitle
\begin{abstract}
  Spatial arbitrage pricing theory (S-APT) extends factor models to allow
returns to spill over between connected firms, beyond what common risk
factors explain. In recent years, tabular foundation models (TFMs) have
matched or outperformed tuned gradient-boosted and bagged ensembles on
small and medium-sized tabular benchmarks. Although their predictive
performance is strong, these models reveal little about what they have
learned, which limits their use in regulated financial applications that
require explanation. Internal representations are widely used to probe
what neural networks learn and to measure similarity. In this paper, we use
the frozen representations of four TFMs as a measure of financial
dissimilarity in a spatial asset-pricing model of US equities from 2011 to
2025, and find that their estimated gains over a direct distance between
the same disclosures become consistently positive once broad SEC filing
data are included. This reframes tabular foundation models from black-box
predictors into instruments for measuring economic structure, with immediate
applications to peer selection, risk attribution and network-based
asset pricing.
More broadly, our findings suggest that the internal representations of
foundation models, not only their predictions, deserve study as economic
objects, opening a new line of research on what pretrained models know
about firms.

\end{abstract}
\ffResultStatus
\section{Introduction}

Spatial arbitrage pricing theory (S-APT) extends the factor structure of
asset returns by allowing shocks to one firm to spill over to the firms
connected to it, so that returns co-move beyond what common risk factors
explain \citep{kou2018}. Because these spillovers travel along a network,
the theory is only as informative as the measure of proximity that defines
it. That measure determines whose returns enter each firm's peer average,
and through the spatial multiplier it also shapes the covariance the model
implies across firms.

Tabular foundation models (TFMs) offer a new source of such a measure.
Pretrained across many datasets, these models now match or outperform tuned
gradient-boosted and bagged ensembles on small and medium-sized tabular
benchmarks \citep{hollmann2025,ye2025}, predicting for a new table in a
single forward pass. Although their predictive performance is strong, they
reveal little about what they have learned, and that opacity limits their
use in regulated financial applications, where model risk management
requires predictions to be explained.

In language, vision and multimodal models, the same opacity has been
approached through the models' internal representations. These
representations are widely used to probe what a network has learned
\citep{kornblith2019,huh2024}, and distances between them serve as
similarity measures for retrieval, zero-shot classification and clustering.
Applied to financial records, a TFM's representation therefore offers both
a window into what the model has learned about firms and a candidate
measure of which firms are close.

Such a representation could in principle be learned directly from
financial data, but statistical learning theory explains why financial
panels make this difficult. When a representation is selected from a
hypothesis set $\mathcal{H}$ by fitting the data, the gap between its
in-sample and population performance is controlled by the capacity of
$\mathcal{H}$, measured by its Vapnik--Chervonenkis dimension or Rademacher
complexity, relative to the number of observations
\citep{shalevshwartz2014}. Because financial
disclosures are low-frequency, our panel of at most 1,000 issuers at each
of fifteen annual formations supplies at most 15,000 firm-years, and
these observations are
dependent across firms and over time. A representation flexible enough to
compare many filing fields belongs to a hypothesis set whose capacity such
a sample cannot discipline. While heavily parameterized models can still
predict returns well out of sample \citep{kelly2024virtue}, their
credibility then rests on held-out evaluation, which an in-sample
comparison of peer graphs does not provide. A frozen pretrained model
removes the estimation step instead. Since its weights are fixed before it
sees the panel, and the labelled historical examples enter only as
conditioning context, no parameter is fitted to the returns on which the
peers are evaluated, and pretraining across many datasets takes the place
of estimation as a prior over prediction tasks
\citep{muller2022,nagler2023}. The only selection made on the panel is
among twelve combinations of encoder and input set, a finite hypothesis set
whose cost grows only with the logarithm of its size.

In this paper, we use the frozen internal representations of four TFMs as a
measure of financial dissimilarity in a spatial asset-pricing model, and
we compare the resulting peers with the characteristic-history distance of
\citet{fernandez2011} and with a direct distance between the same
standardized disclosures. The comparison matters because two firms with
similar leverage can differ widely in cash flow or market exposure, so the
firms that count as peers depend both on the characteristics compared and
on the distance assigned to their differences. Whereas
\citet{fernandez2011} builds peer weights from the co-movement of
characteristic histories, a TFM encodes each firm's financial record
alongside labelled historical examples, and each firm's peers are the firms
whose representations lie closest.

How weight is spread across firms matters as much as the distance itself.
Because the joint characteristic kernel in our US reimplementation of
Fern\'andez's spatial capital asset pricing model (CAPM) gives each issuer
about \ffnum[0]{quarterly-fernandez-joint-effective-neighbours} effective
neighbours among \ffnum[0]{quarterly-primary-issuers} fitted issuers, every
firm's peer return lies close to the cross-sectional average, and the
spatial coefficient can absorb the common mean instead of measuring
co-movement among similar firms. We therefore compare peer graphs on a
quarterly panel in which quarter effects absorb the common return level and
each issuer selects ten neighbours, and we confirm that the ranking of
distances persists under two smooth kernels. Throughout, fit refers to the
in-sample spatial-lag log likelihood per issuer-quarter. Since the encoders
were released after the sample and the filing values include later
revisions, the comparisons measure contemporaneous co-movement
retrospectively rather than forecast returns.

With the peer weights concentrated, the design separates two sources of
improvement that a single benchmark would conflate. On four financial
ratios, all four encoders we study have a higher estimated fit than
Fern\'andez's distance, and for TabPFN the difference survives the five
Fama--French factors and momentum. A second benchmark, which selects peers
by the distance between the same standardized ratios without an encoder,
improves on Fern\'andez's distance by a comparable amount, so that
improvement does not require encoding. Whereas the encoders disagree in
sign against this direct graph on the ratios alone, encoding adds fit once
the ratios are augmented with financial fields from SEC filings: all four
encoders then have positive estimated gains over the direct graph built
from the same fields, and the largest, from TabPFN, stays positive under
multifactor controls.

These comparisons establish when an encoder improves fit, although neither
fit nor prediction accuracy reveals how the encoder organizes firms.
Because an invertible recoding can preserve every prediction while
reordering distances, Proposition~\ref{prop:recovery} shows that
representations equivalent for prediction can select different peers, and a
four-firm example shows the reordering changing a symmetrized neighbour
graph. Proposition~\ref{prop:multiplier} then bounds how far such a change
in peer weights can move the covariance implied by the spatial model. The
selected broad-input TabPFN representation exhibits this gap in the data.
Although its leading principal coordinate carries \ffmatched{inertia}\% of
positive inertia and orders firms by market equity over assets, a ratio the
model receives both as an input and as the label of its historical
examples, peers chosen along that coordinate have a mean Jaccard similarity
of only \ffmatched{overlap} with the full representation's peers, and
restricting the graph to it lowers fit by more than the encoder's entire
gain over the direct graph. While the valuation axis summarizes the
representation, the peers that improve fit depend on the dimensions beyond
it.

Taken together, these results make three contributions. The first is to
use the frozen internal representations of TFMs as the proximity measure in
a spatial asset-pricing model, evaluated against a matched direct
benchmark that separates the choice of distance from the use of an
encoder. The second is evidence on how that comparison varies across four
TFMs, three input sets and nested factor controls. The third is an
account, formal and empirical, of what probing such a representation
reveals, and of why neither prediction accuracy, nor return fit, nor the
leading axis identifies the peers it selects.

\section{Related Work}

Several literatures define peers through economic links and show that the
returns of linked firms carry information about one another, with customer
and supplier relationships \citep{cohen2008}, technological proximity
\citep{lee2019technological} and the similarity of product descriptions
\citep{hoberg2016} each identifying a different kind of connection. Whereas
these approaches start from a named relationship, \citet{fernandez2011}
uses the co-movement of firms' financial characteristics over time to
define distances and converts those distances into weights with an
exponential kernel. Spatial arbitrage pricing theory (S-APT) shows how such
a network, once fixed, restricts expected returns when shocks spill over
between connected firms \citep{kou2018}, so each of these measures is a
candidate for the proximity that S-APT takes as given. Because the
economic object in our
application is financial co-movement, we evaluate a set of peers by the
in-sample association of issuer returns with contemporaneous peer returns
after factor adjustment, using the spatial lag and error models of
\citet{anselin1988} and \citet{lesage2009}.

Representation learning in finance has mostly learned firm structure from
returns or holdings: \citet{kelly2019} treat characteristics as
instruments for time-varying factor exposures, \citet{gu2021} learn how
exposures vary with characteristics in an autoencoder asset-pricing model,
and \citet{gabaix2025asset} embed assets through investor holdings.
Whereas these models estimate their representations from financial data,
a tabular foundation model (TFM) arrives with a representation fixed before
it sees any financial record, because it is pretrained across many
unrelated datasets. In a prior-data fitted network, pretraining samples
tasks and teaches the predictor to infer labels given examples from a new
table \citep{muller2022,nagler2023}, and \citet{hollmann2025} show that
TabPFN supports both prediction and reusable embeddings. Since benchmark
evidence for TFMs comes mostly from independent and identically
distributed tables, \citet{purucker2026} ask how far their performance
extends beyond that setting, a question that financial panels, dependent
over time and across firms, make pressing. In our application the
historical examples carry observed next-quarter valuation labels, and we
select peers by the distances between the resulting issuer vectors.

Because identification results for learned representations characterize
them only up to a class of transformations, the choice among equivalent
representations can matter for peer selection. \citet{roeder2021}
establish identification up to an invertible linear transformation for a
family of discriminative models, and identifiability results for
latent-variable models likewise hold only up to an equivalence class
\citep{khemakhem2020}. Although common indices of representation
similarity are invariant to orthogonal transformations and isotropic
scaling \citep{kornblith2019}, which preserve the ordering of distances, a
general invertible linear map can stretch one direction more than another
and thereby reorder neighbours. \citet{steck2024} make the corresponding
point for embeddings, showing that cosine similarities can be arbitrary
when a model leaves the scale of each embedding dimension free, so the
similarities used to probe a representation can depend on coordinates
that its predictions leave undetermined. Our four-firm construction makes
this concrete for peer selection, since a change of coordinates preserves the
recoverable predictions while changing the peer weights in a spatial model.

\section{Pretraining and Financial Proximity}

Although a predictor and a spatial model use the same issuer vector, they
use it in different ways, since the predictor reads out a quantity from
each vector whereas the spatial model compares vectors across firms.
When a TFM's internal representation serves as a similarity measure, the
question is therefore whether the predictions it supports also determine
which firms are close.
For a fixed dated context, let $F(h)$
denote the ideal task prediction for history $h$ and $s(h)$ the economic
target of interest. An exact readout is a function of the representation
that reproduces the specified quantity for every history.

\begin{proposition}[Task recovery and peer selection]
  \label{prop:recovery}
  Suppose the target space is nonempty. Every representation that admits
  an exact readout of $F$ admits an unrestricted readout of $s$ if and only
  if $F(h)=F(h')$ implies $s(h)=s(h')$ for all histories.
  For any representation $z$ with $s=r\circ z$ and any bijection $g$ of
  its representation space, $r\circ g^{-1}$ recovers $s$ from $g\circ z$.
  Such a recoding can change the symmetrized nearest-neighbour graph.
  \lean{FreeFoundations.Alignment.optimal_imp_sufficient_iff}
\end{proposition}

The first statement follows by taking $F$ itself as the representation
in the forward direction. In the reverse direction, each value of $F$
determines one value of $s$, so a readout of $F$ also supplies a readout
of $s$. The second statement matters whenever the return model assigns
peer weights from distances between the vectors.

Consider $A=(1,1)$, $B=(1,0)$, $C=(1,4)$ and $D=(1,5)$.
Stretch the first coordinate by four. Dividing that coordinate by four
before applying the original readout preserves every prediction.
For peer selection, normalize each vector to unit
length and compare chord distances. The closest peers of $A,B,C,D$ are
initially $C,A,D,C$ and become $B,A,D,C$. Joining firms whenever either
selects the other therefore removes the $AC$ edge and changes peer weights.
\lean{FreeFoundations.Neighbours.symmetrized_chord_edge_flip}

The example uses one neighbour to make the edge change visible. Our
main panel comparison uses ten neighbours per issuer before symmetrization.

\paragraph{Distances and weights.}
For the US benchmark, we compute Spearman chord distance for each
characteristic over paired annual observations. For a pair with
Spearman correlation $r$, this distance is $\sqrt{2(1-r)}$. Pairs
with fewer than
two common years or an undefined correlation receive the neutral
distance $\sqrt{2}$. The joint distance is the root mean square
across the four characteristics. We set the diagonal of $K$ to zero,
set $K_{ij}=\exp(-d_{ij})$ off the diagonal, and divide each row by
its sum to obtain the weight matrix $W$. The effective number of
neighbours for issuer $i$ is
$1/\sum_j W_{ij}^2$, the number of equally weighted peers that would
give the same weight concentration. On normalized nonzero embeddings
the chord distance is the
Euclidean distance between unit vectors. To construct a binary
$k$-nearest-neighbour graph we join two firms if either selects the
other before row normalization. Both constructions use a symmetric raw affinity.

\paragraph{Spatial specification.}
In the fixed-cohort cross-section, write $y_i$ for an issuer's median
quarterly excess return and $b_i$ for its estimated CAPM beta. We compare
the conventional cross-section with the spatial lag model
$y=\alpha\mathbf{1}+\gamma b+\rho Wy+u$
\citep{anselin1988,lesage2009}, with $\alpha$ either fixed to zero or
estimated. The corresponding spatial-error specification
applies the multiplier to the error and leaves the systematic term
unchanged. With $\alpha=0$, the peer average may absorb the
cross-sectional mean. With an intercept, exact uniform off-diagonal
weights make the added regressor a linear combination of a constant
and $y$, so the spatial term requires variation beyond the sample
mean and the mechanical own-observation correction.

Peer weights also enter the spatial multiplier.
Proposition~\ref{prop:multiplier}
bounds the resulting change in implied covariance for a common roster,
spatial coefficient and innovation covariance.

\paragraph{Estimation and uncertainty.}
For the characteristic benchmark the estimated quantities include the
market betas, the cross-sectional premium, the intercept when
included, the spatial coefficient, and the error variance. For the
encoder comparisons, we fix the model weights, the rule for
selecting historical examples, the preprocessing and the distance measure,
so the only quantities estimated on the panel are the factor exposures and
the parameters of the return model.
For each matched-panel contrast, we divide the difference in total log
likelihood by the number of matched issuer-quarters. We use natural
logarithms, so this difference is in nats per issuer-quarter, and we
refer to it as the difference in fit. A positive value favours the first
graph. We draw 9,999 stationary bootstrap samples of
quarters, using the same time blocks for both fits \citep{politis1994}.
Within each draw we divide the summed difference in quarterly
contributions by the resampled issuer-quarter count. The central 95\%
of these draws describes time variation conditional on the fitted
loadings, model parameters and graphs. In the fixed-cohort exercise,
where issuer betas and median returns
enter a single cross-section, we recompute both quantities in each refit.

\section{Empirical Design and Data}

\begin{table*}[t]
  \centering
  \fftabletype
  \IfFileExists{generated/tables/matched-factors.tex}{\begin{tabular}{lllll}
\toprule
Control & Baseline LL & Fernández gain & TabPFN gain & TabPFN minus Fernández \\
\midrule
Market & 27,996.4 & 0.0024 [0.0010, 0.0041] & 0.0074 [0.0030, 0.0125] & 0.0050 [0.0015, 0.0088] \\
Momentum & 27,136.7 & 0.0027 [0.0011, 0.0045] & 0.0078 [0.0036, 0.0127] & 0.0051 [0.0017, 0.0088] \\
Market + Mom. & 28,785.7 & 0.0022 [0.0008, 0.0036] & 0.0071 [0.0030, 0.0117] & 0.0049 [0.0016, 0.0085] \\
FF5 & 31,968.2 & 0.0007 [0.0003, 0.0012] & 0.0019 [0.0010, 0.0029] & 0.0012 [0.0003, 0.0021] \\
FF5 + Mom. & 32,148.2 & 0.0008 [0.0003, 0.0014] & 0.0021 [0.0011, 0.0032] & 0.0013 [0.0003, 0.0024] \\
\bottomrule
\end{tabular}
}{\ffmissing}
  \caption{Baseline LL is the total factor-only log likelihood. The next
    two columns give gains over that baseline in nats per issuer-quarter.
    The final column compares the two peer graphs in the same units.
    Brackets give 95\% intervals from resampling paired quarterly
  contributions at the fitted loadings and parameters.}
  \label{tab:matched-factors}
\end{table*}

\begin{table*}[t]
  \centering
  \fftabletype
  \IfFileExists{generated/tables/matched-encoders.tex}{\begin{tabular}{llll}
\toprule
Encoder & Ratio histories & Underlying financials & Broad financials \\
\midrule
TabPFN 3.5 & -0.0019 [-0.0040, 0.0002] & 0.0040 [-0.0002, 0.0091] & 0.0085 [0.0032, 0.0154] \\
TabFM & 0.0022 [0.0002, 0.0045] & -0.0039 [-0.0054, -0.0025] & 0.0047 [0.0008, 0.0096] \\
Kumo & 0.0018 [-0.0006, 0.0046] & -0.0050 [-0.0068, -0.0033] & 0.0034 [-0.0008, 0.0084] \\
LimiX & -0.0032 [-0.0056, -0.0010] & 0.0011 [-0.0015, 0.0041] & 0.0054 [0.0009, 0.0110] \\
\bottomrule
\end{tabular}
}{\ffmissing}
  \caption{Market-controlled gain in fit of each encoder graph over the
    direct graph built from the same inputs, in nats per issuer-quarter.
    Columns give the ratio, underlying and broad input sets. Brackets give
    a conditional 95\% interval for each comparison. All graphs use
  symmetrized ten-neighbour weights.}
  \label{tab:matched-encoders}
\end{table*}

Because the diagnostic of characteristic weights and the comparison of
peer graphs ask different questions, we use two samples. The
cross-sectional spatial CAPM relates each
issuer's median quarterly excess return to its market exposure within a
single formation cohort. The panel comparison follows successive December
rosters, each formed from the 1,000 largest US common stocks and grouped by
company into issuer portfolios, through the four quarters of the following
year, covering 2011--2025. Each January we encode the issuers on the
preceding December
roster. In each quarter we use those representations to select peers
among the issuers with eligible returns and factor exposures.

For the single cohort, we select the 1,000 largest US common securities
by December 2015 market capitalization in CRSP CIZ. Grouping share classes
(PERMNOs) by company (PERMCO) gives
\ffnum[0]{quarterly-primary-population-issuers} issuer portfolios.
We follow the selected classes through 2016--2025, including those that
fall below the ranking cutoff. Linkage and feature requirements leave
\ffnum[0]{quarterly-primary-roster} issuers on the analytical roster.
Of these, \ffnum[0]{quarterly-primary-issuers} have at least eight
observed quarters and a converged robust regression estimate of market
exposure. Appendix~\ref{app:data} details the treatment of incomplete
quarters and delistings.

For the panel we compare three input sets, which we call the ratio,
underlying and broad inputs. The ratio inputs contain ten annual slots for
each of four ratios: market equity over assets, market-to-book, dividend
yield and debt maturity. The underlying inputs contain histories of the
numerators and denominators used to construct those ratios. The broad
inputs augment the ratio histories with the standard-taxonomy financial
fields from SEC CompanyFacts that are reported for at least 10\% of
issuer-quarters in the financial panel. We study four TFMs, TabPFN 3.5
(building on \citealp{hollmann2025}), TabFM 1.1 \citep{kong2026tabfm},
Kumo-Tabular \citep{nvidia2026kumo} and LimiX-2 \citep{zhang2026limix},
and call each an encoder when it produces issuer representations
(Appendix~\ref{app:data}). Their weights are never updated. At each January
formation, the frozen encoder receives up to 256 historical examples
labelled with an already observed value of next-quarter market equity over
assets, which condition the representation without fitting any
parameter.
For wide tables we encode successive blocks of up to 500 columns and
concatenate the issuer vectors. Filing cutoffs date the inputs, but the
filing values include later revisions and the encoders were released after
the sample, so the exercise is retrospective.

Three graphs select peers among the same issuers. The Fern\'andez graph
uses the joint distance, which compares the ranks of paired annual
observations within each characteristic and is available only for the ratio
histories. The direct graph uses chord distances between preprocessed
financial feature vectors, for which we impute missing features with
historical-example medians, retain missingness indicators and standardize
each feature, so every standardized field receives the same weight. The
encoder graph uses chord distances between the issuer representations that
a frozen encoder produces from the same features together with the
labelled historical examples. For each input set the encoder and direct
graphs share financial fields, issuer-quarters and neighbour rule, so their
difference in fit compares two complete peer-selection methods, including
the encoder's use of the labels.

Let $y_t$ be the vector of observed issuer excess returns in quarter $t$.
We fit the spatial lag model
\[
  y_t=\alpha_t\mathbf{1}+X_t\gamma+\rho W_t y_t+u_t .
\]
The vector $W_t y_t$ contains contemporaneous peer returns. Adding this
spatial lag to a factor model gives the return structure from which S-APT
derives its pricing restrictions \citep{kou2018}, and we estimate that
structure under each peer graph without imposing the restrictions. Quarter
effects absorb the common return level. Each column of $X_t$ multiplies
an issuer's estimated exposure by the corresponding quarterly factor
return. We consider individual factors and joint controls including the
five Fama--French factors (FF5) plus momentum
\citep{fama2015,carhart1997}. For the nonmarket factors we compound the
monthly long--short returns from the French data library as quarterly
proxies. We estimate issuer exposures over each issuer's full observed
quarterly panel, separately for each factor specification. All matched fits
use issuer-quarters with observed returns and converged exposures across
the factor specifications. We compare the maximized Gaussian log likelihoods
on this common sample.

\section{Results}

\begin{table*}[t]
  \centering
  \fftabletype
  \IfFileExists{generated/tables/matched-ablation.tex}{\begin{tabular}{lll}
\toprule
Peer graph & Mean overlap & Fit difference [95\% interval] \\
\midrule
First coordinate & 0.053 & -0.01157 [-0.01682, -0.00700] \\
First three coordinates & 0.218 & -0.00531 [-0.00793, -0.00302] \\
Omit market equity / assets & 0.250 & 0.00073 [-0.00083, 0.00221] \\
Omit market-to-book & 0.792 & 0.00060 [-0.00001, 0.00129] \\
Omit dividend yield & 0.809 & 0.00038 [-0.00037, 0.00115] \\
Omit debt maturity & 0.808 & 0.00031 [-0.00040, 0.00104] \\
\bottomrule
\end{tabular}
}{\ffmissing}
  \caption{Broad-input TabPFN peer graphs after projection onto leading
    principal coordinates or omission of one engineered ratio history.
    Overlap averages issuer-level Jaccard similarities with the full graph
    within each quarter and then across quarters. Fit differences are
    market-controlled, in nats per issuer-quarter relative to the full
  graph, with paired conditional 95\% intervals.}
  \label{tab:ablation}
\end{table*}

\paragraph{Concentration of characteristic weights.}
Because a comparison of distances is informative only when peer weights
differ across firms, we begin with the fixed cohort, where the
characteristic weights turn out to be nearly uniform. In the December 2015
cohort, the joint exponential kernel gives each issuer an average of
\ffnum[0]{quarterly-fernandez-joint-effective-neighbours} effective
neighbours among the \ffnum[0]{quarterly-primary-issuers} fitted issuers,
so the weight is spread across nearly the whole roster and the peer
average varies little from one issuer to the next. With the intercept
fixed at zero the estimated spatial coefficient is
$\widehat{\rho}=\ffnum{quarterly-fernandez-joint-rho}$, whereas allowing an
intercept produces an invalid or boundary fit
(Table~\ref{tab:us-benchmark}). Because the peer average is nearly common
to all issuers, the large coefficient is consistent with the peer return
absorbing the cross-sectional mean, and it says little about which firms
the kernel treats as close. The repeated-formation panel addresses both
difficulties, since quarter effects absorb the common return level and a
ten-neighbour rule concentrates each issuer's peer return on a
small group whose composition depends on the distance.

\paragraph{Peer returns after factor adjustment.}
Once quarter effects and concentrated weights are in place, both distances
produce peer returns that add to a factor-only model, and the advantage of
TabPFN over Fern\'andez's distance survives the full set of factor
controls. Across \ffmatched{observations}
matched issuer-quarters, adding the peer return from either the Fern\'andez
graph or the ratio-based TabPFN graph raises fit relative to the
factor-only model under every control set in
Table~\ref{tab:matched-factors}. Although the difference between the two
graphs narrows as the controls expand, almost all of the narrowing occurs
when the size, value, profitability and investment factors enter,
consistent with peers selected on valuation and financing ratios sharing
exposure to those factors, and after FF5 plus momentum the TabPFN graph
still fits better, by \ffmatched{factor-gain} nats per issuer-quarter (paired
95\% interval \ffmatched{factor-interval}).

\begin{figure}[t]
  \centering
  \IfFileExists{generated/figures/matched-kernels.pdf}{\includegraphics[width=\columnwidth]{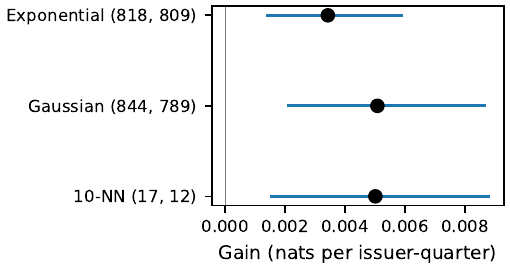}}{\ffmissing}
  \caption{Market-controlled gain in fit of the ratio-based TabPFN graph
    over the Fernández graph under each weighting rule, with conditional
    paired 95\% intervals. Parentheses give mean effective neighbours for
  Fernández and TabPFN, respectively.}
  \label{fig:matched-kernels}
\end{figure}

While the Fern\'andez graph draws on all four characteristics, the
coverage of their histories limits it early in the panel. Because a rank
correlation requires at least two overlapping annual observations, and in
the first panel quarter nearly all dividend-yield pairs lack them, those
pairs receive the same neutral distance, and the joint distance averages in
a characteristic that cannot separate firms (Table~\ref{tab:neutral}).
This ranking of the two graphs does not depend on the ten-neighbour rule,
since under market controls the ratio-based TabPFN graph also fits
better than the Fern\'andez graph when both use
exponential weights or both use median-bandwidth Gaussian weights
(Figure~\ref{fig:matched-kernels}), even though the smooth kernels spread
weight across hundreds of effective neighbours rather than concentrating it
on ten.

\paragraph{Where the improvement over characteristic histories arises.}
Since the comparison with Fern\'andez's distance changes two things at
once, namely the way the ratio histories are compared and whether they pass
through an encoder, the direct graph is needed to separate them, because it
compares the same standardized ratio histories that TabPFN receives
without encoding them. On the quarterly panel the direct graph
improves on the Fern\'andez graph by
more than the ratio-based TabPFN graph does, at 5.91 against 4.98
log-likelihood units per quarter, and the paired difference between TabPFN
and the direct graph has an interval spanning zero
(Table~\ref{tab:quarterly-comparisons}). All four encoders improve on the
Fern\'andez graph on the ratio inputs, by 3.30 to 8.09 units per quarter
with intervals above zero, so the direct graph's gain lies within the range
the encoders achieve. The same ordering holds on the matched factor
sample, where the estimated gain of TabPFN over the direct
graph on the ratio inputs is slightly negative and its 95\% interval ends
at 0.0002 nats per issuer-quarter, a small fraction of the TabPFN graph's
advantage over the Fern\'andez graph (Tables~\ref{tab:matched-factors}
and~\ref{tab:matched-encoders}). Because the direct graph differs from the
Fern\'andez graph in several respects at once, comparing the levels of
the standardized histories rather than their rank co-movement, imputing
missing values rather than assigning a neutral distance, and standardizing
each field, this decomposition shows that the improvement over
Fern\'andez's distance does not require an encoder, while leaving open
which of those differences produces it. Because a size-based graph
and a permutation null both fit worse than the Fern\'andez graph, the
characteristic-history distance does carry information about peers, which
the direct comparison improves on.

\paragraph{When encoding adds fit.}
Having established that the improvement over Fern\'andez's distance does
not require an encoder, we turn to whether an encoder ever improves on the
direct graph, and the four encoders agree in sign against it only on the
broad inputs (Table~\ref{tab:matched-encoders}). On the ratio inputs they
scatter around the direct graph, with one interval above zero and one
below, and on the underlying inputs two of the four fit worse than the
direct graph. On the broad inputs, by contrast, all four estimated gains
are positive and three of the four intervals exclude zero, with TabPFN
recording the
largest, at \ffmatched{broad-gain} nats per issuer-quarter. Because the
direct distance weights every standardized field equally, adding many
filing fields reduces the relative weight of the four ratios, and on the
quarterly panel the direct graph's advantage over the Fern\'andez graph is
smaller with the broad inputs than with the ratios alone
(Table~\ref{tab:quarterly-comparisons}). The encoders instead
condition their representations on labelled historical examples, and with
the broad inputs each of them improves on the direct graph in point
estimate.

Although the broad-input TabPFN gain is smaller under FF5 and FF5 plus
momentum than under market controls, mirroring the narrowing in
Table~\ref{tab:matched-factors}, its conditional interval remains above
zero in each specification (Table~\ref{tab:matched-broad-factors}). We
select this representation for geometric analysis by its market-controlled
rank among the twelve estimates, and since each interval describes its own
contrast with the direct graph, none adjusts for that selection.

\paragraph{What the leading coordinate measures.}
Once a representation improves fit, the natural next question is which
characteristics organize it, and the answer is clearest where the inputs
are fewest. In the January 2016 ratio-only TabPFN embedding, the first
three principal coordinates explain \ffmatched{historical-value-r2}\% of
the rank variation in market equity over assets among
\ffmatched{historical-value-n} issuers but only
\ffmatched{historical-beta-r2}\% of the rank variation in market beta among
the \ffmatched{historical-beta-n} issuers with return-based estimates
(Appendix~\ref{app:scope}). Although close alignment with valuation is
expected, because valuation is both an input and the label of the
historical examples, market beta, which the encoder never receives, is the
more informative check, and the coordinates carry almost none of its
variation. The selected broad-input representation shows the same
organization: in a principal-coordinates decomposition of its January
2016 distances, the leading eigenvalue
accounts for \ffmatched{inertia}\% of the sum of positive eigenvalues,
above the 25\% threshold we fixed in advance for analysing a single axis,
and that axis orders issuers by market equity over assets with an absolute
rank correlation of \ffmatched{valuation}.

\paragraph{Peer sets beyond the leading coordinate.}
While a dominant valuation axis makes the representation easy to
summarize, peer selection depends on which firms lie close in every
direction. At each
annual formation we project the issuers onto the leading coordinate and
compare each firm's neighbours along that axis with its neighbours in the
full representation. Averaged first across firms and then across quarters,
the two peer sets have a Jaccard similarity of \ffmatched{overlap}, and
restricting the graph to the axis changes fit by
$\ffmatched{projected-gain}$ nats per issuer-quarter under market controls
(Table~\ref{tab:ablation}). Since this loss exceeds the encoder's
entire estimated gain over the direct graph, and retaining three
coordinates raises the overlap and recovers only part of the loss, the
valuation axis summarizes the representation without reproducing its
peers, and the fit that the encoder adds over the
direct graph depends on the dimensions beyond it.

These projections remove dimensions from the representation, whereas
omitting each engineered ratio history in turn removes information from its
inputs, and the omissions show the converse pattern. Although removing the
market-equity-to-assets history, with the broad filing fields and the
valuation labels still available, lowers peer overlap with the full graph
to a Jaccard similarity of 0.25, its interval for the change in fit,
$[-0.0008, 0.0022]$ nats per issuer-quarter, excludes any loss as large as
the projection's, and the other three omissions leave most peers in place.
Because similar fit can therefore accompany substantially different peer
sets, fit does not determine a representation's peers any more than
prediction accuracy does in Proposition~\ref{prop:recovery}
(Appendix~\ref{app:ablation}).

\section{Conclusion}

Spatial arbitrage pricing theory takes a measure of proximity between firms
as given, and our results show that the frozen internal representations of
TFMs can supply one, although what they add depends on the benchmark
against which they are judged. Using four financial ratios, every TFM we
study has a higher estimated fit than Fern\'andez's characteristic-history
distance, yet a direct distance on the same standardized ratios improves
on that distance by a comparable amount, so the improvement does not
require an encoder. Once broad filing data are added, by contrast, all four
TFMs have positive estimated gains over the direct distance, and the
largest survives five Fama--French factors and momentum.

Probing these representations shows what they have learned about firms.
In the ratio-only embedding, the leading coordinates explain almost all of
the rank variation in valuation and almost none of the variation in market
beta, and the selected broad-input representation is organized along the
same valuation axis. That dominant axis does not determine the peers,
however. Firms close on it have few neighbours in common with the full
representation, and selecting peers along it lowers fit by more than the
model's gain over the direct distance. Omitting the valuation history
changes peers substantially, yet its interval for the change in fit
excludes a loss of that size. Read alongside
Proposition~\ref{prop:recovery}, these results show that neither
prediction accuracy, nor return fit, nor the leading axis identifies the
peers a representation selects.

An encoder proposed as the proximity measure for S-APT should therefore be
compared with a direct distance on its own inputs, and its peers should be
read from the full representation rather than from its dominant
coordinate. Because our comparisons are in-sample and retrospective, the
natural next step is held-out evaluation at each formation date, using the
model releases and filing values then available. Such an evaluation would
also allow a fair comparison with representations trained on financial
data, whose capacity held-out performance can discipline
\citep{kelly2024virtue}.

\ffbibliography{plainnat}
\appendix
\section{Technical Proofs and Formalization}
\label{app:proofs}

\paragraph{Economic distance and task recovery.}
For loadings $b_i,b_j$ on mean-zero common shocks $q$ with finite second
moments, the root mean square of $(b_i-b_j)^\top q$ measures the difference
in return responses in units of return. Writing $\Omega$ for the shock
covariance and $a_i=\Omega^{1/2}b_i$ gives the distance $\norm{a_i-a_j}$.
\lean{FreeFoundations.Score.response_sq}
Perfect marginal forecasts can be consistent with different joint responses:
two fair binary returns can agree in one regime and move oppositely in
another while each retains the same marginal distribution.
\lean{FreeFoundations.Alignment.marginals_free_of_state}
\lean{FreeFoundations.Alignment.joint_depends_on_state}

Suppose ideal task predictions satisfy $\tau=Ca+v$, where $a$ is a
risk-scaled exposure and $v$ the mismatch. For a finite-dimensional
exposure space, a linear map satisfying $\norm{Cu}\geq\kappa\norm u$
in all directions with $\kappa>0$ admits a linear decoder $L$ with
\[
  \norm{L\widehat\tau-a}\leq
  (\norm{\widehat\tau-\tau}+\norm v)/\kappa .
\]
Thus the bound tightens as the tasks distinguish exposure directions more
sharply. For finite labels and contexts with positive predicted
probabilities, excess log loss measures average divergence from the task's
conditional label law.
\lean{FreeFoundations.Recovery.exists_decoder}
\lean{FreeFoundations.Recovery.exposure_error_mismatch}
\lean{FreeFoundations.Score.pretraining_excess}

\paragraph{Recoverability and the transformation class.}
If $s=r\circ z$, histories in the same fibre of $z$ share the same
target. Conversely, define $r$ on $z(H)$ by the unique target of each
fibre and use any value in $S$ outside the image. In the chord example
the four vectors have coordinates $(x,y)$ with $y=1,0,4,5$ respectively.
The stretch takes $x=1$ to $x=4$. For heights $y,z$ the squared cosine is
\[
  c^2=\frac{(x^2+yz)^2}{(x^2+y^2)(x^2+z^2)}.
\]
All dot products are positive, so the unit-vector chord is
$\sqrt{2-2\sqrt{c^2}}$. For $AB$ the squared cosine rises from $1/2$
to $16/17$, while for $AC$ it stays at $25/34$. This reverses the
first firm's choice even on the three-firm roster. With the fourth
firm present, $C$ and $D$ select one another before and after the stretch.
\lean{FreeFoundations.Neighbours.chord_neighbour_flip}
Composing the raw-state readout with the inverse stretch preserves
predictions. For a
restricted readout class $\mathcal R$, a bijection $g$ preserves
the class when $\{r\circ g^{-1}:r\in\mathcal R\}=\mathcal R$. A
recoding with this property leaves the attainable predictions unchanged.
The bijections that preserve $\mathcal R$ form a group under
composition. The fibre
argument is pointwise. A measurable or approximate recovery claim
requires additional assumptions on the spaces and readout class.

Although recoding can change peers, an estimated exposure can
preserve the selected peers when the gap at the neighbourhood
boundary is sufficiently wide. Suppose $a_i$ is a true exposure,
$\widehat a_i$ its estimate, and
$\norm{\widehat a_j-a_j}\leq\delta_j\leq\delta_{\max}$ for every
issuer. Each distance in row $i$ can move by at most
$\delta_i+\delta_{\max}$. Therefore an estimated strict gap between
selected and unselected neighbours exceeding
$2(\delta_i+\delta_{\max})$ preserves that set. For the study's
ten-neighbour rule the relevant gap is between the tenth
and eleventh distances. The certificate applies to estimated exposure
distances when the exposure-error bounds are available.

\paragraph{Statistical alignment and transfer.}
Under squared loss the economic conditional mean must be a measurable
function of the task conditional mean almost surely for every representation
attaining the latter to attain the former.
\lean{FreeFoundations.Sufficiency.sufficient_iff}
Transporting a source-law task error to issuers requires further premises.
If the issuer law is bounded by $\omega$ times the source law, source
root-mean-square error enters the recovery bound multiplied by
$\sqrt{\omega}$. Changes in conditional task means and task--exposure
mismatch enter separately, all divided by the positive separation margin.
\lean{FreeFoundations.Transfer.source_target_recovery}

\paragraph{Multiplier and covariance.}
\begin{proposition}[Spatial stability]
  \label{prop:multiplier}
  Let $K=K^\top\geq 0$ entrywise, with positive degrees $d_i=\sum_j
  K_{ij}$, and set $W=D^{-1}K$ for $D=\operatorname{diag}(d_i)$. For
  $|\rho|<1$, the inverse $P=(I-\rho W)^{-1}$ exists and
  \[
    \norm{P}_2\leq \frac{\sqrt{d_{\max}/d_{\min}}}{1-|\rho|}.
  \]
  On the same roster, let $\widehat W$ have multiplier $\widehat P$
  at the same $\rho$. If both inverse norms are at most $M$ and the
  common innovation covariance $V$ satisfies $\norm{V}_2\leq v$, then
  \[
    \norm{\widehat P V\widehat P^\top-PVP^\top}_2
    \leq 2v|\rho|M^3\norm{\widehat W-W}_2 .
  \]
\end{proposition}
Symmetry makes $d$ stationary for $W$. Jensen's inequality gives
$\norm{Wx}_D\leq\norm{x}_D$, where $\norm{x}_D^2=x^\top Dx$. Hence
$\norm{(I-\rho W)x}_D\geq(1-|\rho|)\norm{x}_D$.
In finite dimension this implies invertibility, and comparing the
weighted and Euclidean norms gives
Proposition~\ref{prop:multiplier}'s first inequality. A symmetric
affinity with isolates gives a block diagonal operator with a zero
block on the isolates, whose multiplier is identity. The bound
applies on the positive-degree block.

The inverse identity gives $\widehat P-P=\rho \widehat P(\widehat
W-W)P$. Expanding the covariance difference as
$(\widehat P-P)V\widehat P^\top+PV(\widehat P-P)^\top$
and using submultiplicativity gives the second inequality. A fixed
portfolio with weights $w$ multiplies the covariance error bound by
$\norm{w}_2^2$. The displayed bound holds with common $\rho$ and $V$.
Changes to those quantities contribute additional terms.

\paragraph{Formalized components.}
The fibre criterion and coding invariance have Lean formalizations.
\lean{FreeFoundations.Alignment.state_optimal_iff}
\lean{FreeFoundations.Geometry.sufficient_equiv}
Lean also checks the neighbour certificate, multiplier bound, and
covariance bound.
\lean{FreeFoundations.Neighbours.neighbourSet_of_estimated_margin}
\lean{FreeFoundations.ReversibleNetwork.normalize_resolvent_l2_bound}
\lean{FreeFoundations.Stability.matrix_covariance_sub_le}
The restricted class group argument and isolate decomposition are
elementary written arguments.

\section{Data Construction and Estimation Details}
\label{app:data}

We weight selected classes by their prior calendar-month
capitalization. CRSP CIZ monthly total returns already incorporate
reported delisting returns. We compound the three observed months of
a quarter once, keeping incomplete quarters unavailable. When a
return row is missing, the absence of a separate terminal-event
flag leaves the cause of exit and its proceeds unknown. The return estimates
therefore describe the observed quarters of the formation cohort.
Formation membership is fixed at the December ranking cutoff. An
issuer retains its formation identity when a selected security's
subsequent source issuer changes. Missing returns and missing
prior-month capitalization are explicit. SEC information uses filing
timestamps before the preceding CRSP close. Negative book equity
where the ratio calls for a positive denominator, stale balance
sheets, and undefined paired history correlations each have separate
availability rules.

The characteristic reimplementation estimates OLS, spatial lag by
instrumental variables \citep{kelejian1998} and maximum likelihood
\citep{lee2004}, and spatial error by maximum likelihood. Fit states
distinguish available estimates from
invalid or boundary fits. For the primary formation, the stationary
quarter bootstrap resamples common date blocks, recomputes betas and
medians, and refits models while the formation graph stays fixed
before filtering unavailable issuers and renormalizing its rows. A
bootstrap distribution for a constrained refit can still be computed
when the corresponding original fit is unavailable. In that case the
percentiles describe the constrained refits, while the
original-sample estimate retains its recorded fit status.

\paragraph{Frozen model readouts.}
For TabPFN 3.5 we take the native query embedding of a single regression
ensemble member. For TabFM 1.1 we extract query rows after the in-context
predictor's final layer normalization \citep{kong2026tabfm}. For
Kumo-Tabular we take the input to the in-context prediction head, and
for LimiX-2 the input to the regression decoder
\citep{nvidia2026kumo,zhang2026limix}. When financial inputs span several
column blocks, we concatenate the issuer's vectors in block order before
computing chord distances.

\paragraph{Coverage of characteristic histories.}
A neutral distance enters a characteristic comparison when fewer than two
annual observations overlap or the paired ranks have undefined correlation.
Coverage varies across characteristics and over the panel
(Table~\ref{tab:neutral}). For ten-neighbour selection we break distance
ties in the ordered issuer roster before joining the two directed choices.

\begin{table}[t]
  \centering
  \fftabletype
  \IfFileExists{generated/tables/matched-neutral.tex}{\begin{tabular}{lrrr}
\toprule
Characteristic & 2011 Q1 & 2025 Q4 & Pooled \\
\midrule
Market equity/assets & 78.80 & 0.67 & 13.67 \\
Market-to-book & 86.32 & 11.33 & 26.65 \\
Dividend yield & 99.96 & 54.89 & 64.83 \\
Debt maturity & 92.79 & 47.41 & 57.96 \\
\bottomrule
\end{tabular}
}{\ffmissing}
  \caption{Share of issuer pairs assigned neutral distance (\%) in each
    characteristic. The pooled share counts unordered pairs within quarter
  across the matched panel.}
  \label{tab:neutral}
\end{table}

\section{Scope of the Empirical Evidence}
\label{app:scope}

Across issuers in the fixed cohort, we relate median excess returns to a
characteristic-weighted average of peer returns. On the repeated-formation
panel, we compare the likelihood of the observed quarterly returns under
different peer weights. The January 2016 geometry analysis then returns to
the formation roster to examine which financial characteristics organize
firms within the ratio-based embedding.

\begin{table*}[t]
  \centering
  \fftabletype
  \IfFileExists{generated/tables/quarterly-likelihood.tex}{%
\begin{tabular}{llrrr}
\toprule
Network & Reference & $\Delta\ell$/quarter & 95\% CI & Holm $p$ \\
\midrule
Direct features (ratios) & Fern\'andez joint & 5.91 & [2.46, 9.74] & 0.055 \\
TabPFN 3.5 (ratios) & Fern\'andez joint & 4.98 & [1.74, 8.69] & 0.135 \\
TabPFN 3.5 (ratios) & Direct features (ratios) & -0.94 & [-2.67, 0.75] & 0.852 \\
TabFM 1.1 (ratios) & Fern\'andez joint & 7.93 & [3.79, 12.85] & 0.023 \\
TabFM 1.1 (ratios) & Direct features (ratios) & 2.02 & [0.15, 4.19] & 0.553 \\
Kumo-Tabular (ratios) & Fern\'andez joint & 8.09 & [3.27, 13.61] & 0.074 \\
Kumo-Tabular (ratios) & Direct features (ratios) & 2.17 & [-0.15, 4.78] & 0.852 \\
LimiX-2 (ratios) & Fern\'andez joint & 3.30 & [0.77, 6.19] & 0.281 \\
LimiX-2 (ratios) & Direct features (ratios) & -2.61 & [-4.56, -0.78] & 0.162 \\
Direct features (XBRL) & Fern\'andez joint & 4.61 & [1.66, 7.77] & 0.081 \\
TabPFN 3.5 (XBRL) & Fern\'andez joint & 11.28 & [5.64, 17.84] & 0.014 \\
TabPFN 3.5 (XBRL) & Direct features (XBRL) & 6.67 & [1.93, 12.70] & 0.281 \\
TabFM 1.1 (XBRL) & Fern\'andez joint & 7.88 & [3.69, 12.60] & 0.022 \\
TabFM 1.1 (XBRL) & Direct features (XBRL) & 3.27 & [-0.07, 7.51] & 0.852 \\
Kumo-Tabular (XBRL) & Fern\'andez joint & 6.45 & [2.72, 10.71] & 0.057 \\
Kumo-Tabular (XBRL) & Direct features (XBRL) & 1.84 & [-1.69, 6.07] & 0.852 \\
LimiX-2 (XBRL) & Fern\'andez joint & 8.54 & [3.73, 14.26] & 0.057 \\
LimiX-2 (XBRL) & Direct features (XBRL) & 3.93 & [-0.30, 9.10] & 0.852 \\
Size & Fern\'andez joint & -1.61 & [-2.94, -0.50] & 0.191 \\
Permutation null & Fern\'andez joint & -1.69 & [-3.16, -0.47] & 0.244 \\
\bottomrule
\end{tabular}

  }{\ffmissing}
  \caption{Market-controlled in-sample spatial-lag log-likelihood differences
    per quarter on the quarterly panel.
    All graphs use ten-neighbour weights. Ratios denote the ratio
    inputs and XBRL denotes the broad inputs. Positive values
    favour the network in the first column.
    Percentiles resample paired quarterly contributions conditional on
  fitted parameters. The last column reports Holm-adjusted $p$-values.}
  \label{tab:quarterly-comparisons}
\end{table*}

\begin{table}[t]
  \centering
  \fftabletype
  \begin{tabular}{lcc}
    \toprule
    Intercept & $\widehat\rho$ & Refit percentiles \\
    \midrule
    Excluded & \ffnum{quarterly-fernandez-joint-rho} &
    $[\ffnum{quarterly-fernandez-joint-rho-bootstrap-low},\
    \ffnum{quarterly-fernandez-joint-rho-bootstrap-high}]$
    \\
    Included & Invalid or boundary & --- \\
    \bottomrule
  \end{tabular}
  \caption{Joint characteristic kernel lag fit for the December 2015
    US formation.
    Percentiles are the central 95\% of
    \ffnum[0]{quarterly-fernandez-joint-rho-bootstrap-draws}
    stationary-quarter refits. They describe refit variation
    conditional on the formation graph and the observed-quarter design.
  Each draw uses the issuers with sufficient observed returns for that refit.}
  \label{tab:us-benchmark}
\end{table}
\paragraph{Encoder and direct graphs on the quarterly panel.}
Table~\ref{tab:quarterly-comparisons} reports differences per quarter
across \ffvalue{quarterly-panel-issuer-quarters} issuer-quarters, while
Tables~\ref{tab:matched-factors} and~\ref{tab:matched-encoders} report
differences per issuer-quarter on the smaller common factor sample. With
symmetrized ten-neighbour weights, the ratio-based TabPFN graph gains 4.98
log-likelihood units per quarter over the Fern\'andez graph, while the
direct graph on the same ratio inputs gains 5.91, and the paired
encoder-minus-direct difference is $-0.94$ per quarter with a 95\%
interval of $[-2.67,0.75]$. Across the family of quarterly comparisons,
the Holm-adjusted $p$-values \citep{holm1979} are 0.135 for the TabPFN
contrast with the Fern\'andez graph and 0.055 for the direct contrast.

\paragraph{The characteristic benchmark.}
The joint characteristic kernel has
\ffnum[1]{quarterly-fernandez-joint-effective-neighbours} effective
neighbours per row on average, compared with $n-1$ for a fitted roster of
$n$ issuers under uniform
off-diagonal weights. When most issuers receive similar weights,
firms share much of the same peer return. These averages carry little
cross-sectional variation relative to a more concentrated graph. On
the \ffnum[0]{quarterly-primary-issuers} fitted issuers, the
joint-kernel lag maximum-likelihood estimate without an intercept is
$\widehat{\rho}=\ffnum{quarterly-fernandez-joint-rho}$.
Table~\ref{tab:us-benchmark} records the contrast and its
uncertainty. With an intercept, the joint-kernel fit is flagged as
invalid or on the boundary. The no-intercept estimate is consistent
with the spatial term absorbing much of the common mean in this US sample.

The point fit and a separate profile-likelihood calculation agree on
the saved original-sample betas, issuer medians and reconstructed
graph. The percentile endpoints recompute betas and medians in common
time blocks and restrict the graph when firms drop out of a refit.
They quantify a different source of variation from the standard error
conditional on the generated beta. These diagnostics
characterize the US reimplementation, where the original sample is unavailable.

\paragraph{Historical ratio-only geometry.}
To identify which financial histories organize the learned peers, we
form principal coordinates from chord distances of the January
2016 TabPFN states using only the four ratio histories. We can
construct these distances for all \ffnum[0]{quarterly-primary-roster}
issuers on the analytical roster because the calculation requires
no subsequent returns. For each ratio we use the latest
available annual slot and restrict the comparison to its observed
issuers. The leading coordinate has stronger absolute rank correlation with
market equity over assets and market-to-book than with
dividend yield or debt maturity. The same ordering holds for explained
rank variation after regression on the leading three coordinates and
an intercept. This subspace comparison is invariant to sign changes and
orthogonal rotations.

The valuation ratios are part of the encoder inputs. Their alignment
describes how the model organizes its own inputs on this date.
The historical ratio embedding puts about 45\% of positive inertia
in its leading coordinate. For the market-beta comparison we restrict
these same coordinates to issuers with an available return-based beta
estimate and regress beta ranks on the three coordinates and an intercept.
On this formation date the learned geometry is closely associated
with a valuation measure supplied to the encoder, while its alignment
with independently estimated market exposure is weak.

\section{Selected Broad-Input Geometry}
\label{app:ablation}

For the broad-input model selected by its market-controlled gain, we rebuild
the peer graph using the first one or three principal coordinates at each
formation. We also omit each engineered ratio history in turn from both
the historical examples and the issuers being encoded. The underlying broad
filing fields remain available, so the omission measures the incremental
role of the supplied ratio history. Each comparison uses the same return
observations and the full broad-input graph as its reference.
Table~\ref{tab:ablation} reports the results. Both projected graphs fit
less well than the full graph, with intervals below zero. Among the four
omissions, removing the market-equity-to-assets history changes the peer
sets most, and all four omission intervals span zero.

\section{Matched Factor Specifications}

\paragraph{Broad-input encoder versus direct features.}
Table~\ref{tab:matched-broad-factors} compares TabPFN with the direct graph
built from the same broad financial disclosures. Both graphs use the common
issuer-quarter mask and symmetrized ten-neighbour weights. The historical
contexts and extracted representations remain fixed. Each factor
specification has its own spatial-lag fit. We resample paired quarters
at the fitted loadings and parameters. The market-controlled ranking
determines which encoder enters this comparison.

\begin{table}[t]
  \centering
  \fftabletype
  \IfFileExists{generated/tables/matched-broad-factors.tex}{\begin{tabular}{lrr}
\toprule
Control & Gain & 95\% interval \\
\midrule
Market & 0.0085 & [0.0032, 0.0154] \\
FF5 & 0.0021 & [0.0002, 0.0043] \\
FF5 + Momentum & 0.0026 & [0.0006, 0.0049] \\
\bottomrule
\end{tabular}
}{\ffmissing}
  \caption{Broad-input TabPFN minus the direct graph in spatial-lag
    log likelihood, in nats per issuer-quarter. Positive gains favour
    TabPFN. All three controls use identical issuer-quarters and peer
    graphs, with paired conditional 95\% intervals. Dashes denote
  unavailable fits.}
  \label{tab:matched-broad-factors}
\end{table}

The following tables report the factor-only, spatial-lag and spatial-error
fits on the same issuer-quarters. Reading down each table shows how fit
changes as factors enter separately, alongside the market, or jointly.
Within a spatial specification, comparing the two peer graphs at the same
factor controls isolates the choice of distance. The parameter counts and
information criteria refer to the return fit conditional on the estimated
issuer exposures.
\newcommand{\ffFactorDetail}[2]{
  \begin{table*}[p]
    \centering
    \fftabletype
    \input{generated/tables/matched-detail-#1}
    \caption{#2 on the common issuer-quarter mask. Information criteria
    condition on estimated first-stage exposures.}
\end{table*}}
\IfFileExists{generated/tables/matched-detail-none.tex}{
  
  \begin{table*}[p]
    \centering
    \fftabletype
    \begin{tabular}{lrrrr}
\toprule
Factors & Parameters & Log likelihood & AIC & BIC \\
\midrule
Market & 62 & 27996.40 & -55868.80 & -55319.80 \\
SMB & 62 & 27625.28 & -55126.56 & -54577.57 \\
HML & 62 & 27833.83 & -55543.67 & -54994.67 \\
RMW & 62 & 26928.80 & -53733.61 & -53184.61 \\
CMA & 62 & 27382.55 & -54641.10 & -54092.10 \\
Momentum & 62 & 27136.70 & -54149.41 & -53600.41 \\
Market + SMB & 63 & 29081.58 & -58037.17 & -57479.31 \\
Market + HML & 63 & 29979.07 & -59832.14 & -59274.29 \\
Market + RMW & 63 & 28925.24 & -57724.48 & -57166.63 \\
Market + CMA & 63 & 29268.46 & -58410.93 & -57853.08 \\
Market + Momentum & 63 & 28785.74 & -57445.47 & -56887.62 \\
FF5 & 66 & 31968.21 & -63804.42 & -63220.01 \\
FF5 + Momentum & 67 & 32148.21 & -64162.42 & -63569.15 \\
\bottomrule
\end{tabular}

    \caption{Factor-only models on the common issuer-quarter mask. Information criteria
    condition on estimated first-stage exposures.}
\end{table*}
  
  \begin{table*}[p]
    \centering
    \fftabletype
    \begin{tabular}{llrrrr}
\toprule
Factors & Peer graph & Parameters & Log likelihood & AIC & BIC \\
\midrule
Market & Fernández & 63 & 28121.56 & -56117.12 & -55559.27 \\
Market & TabPFN 3.5 & 63 & 28380.84 & -56635.68 & -56077.83 \\
SMB & Fernández & 63 & 27741.38 & -55356.76 & -54798.91 \\
SMB & TabPFN 3.5 & 63 & 27982.21 & -55838.42 & -55280.57 \\
HML & Fernández & 63 & 27908.52 & -55691.04 & -55133.19 \\
HML & TabPFN 3.5 & 63 & 28013.20 & -55900.39 & -55342.54 \\
RMW & Fernández & 63 & 27083.61 & -54041.21 & -53483.36 \\
RMW & TabPFN 3.5 & 63 & 27341.22 & -54556.44 & -53998.59 \\
CMA & Fernández & 63 & 27508.97 & -54891.94 & -54334.09 \\
CMA & TabPFN 3.5 & 63 & 27661.20 & -55196.40 & -54638.55 \\
Momentum & Fernández & 63 & 27275.93 & -54425.86 & -53868.00 \\
Momentum & TabPFN 3.5 & 63 & 27541.97 & -54957.94 & -54400.09 \\
Market + SMB & Fernández & 64 & 29179.52 & -58231.04 & -57664.33 \\
Market + SMB & TabPFN 3.5 & 64 & 29407.27 & -58686.55 & -58119.84 \\
Market + HML & Fernández & 64 & 30016.55 & -59905.10 & -59338.40 \\
Market + HML & TabPFN 3.5 & 64 & 30082.02 & -60036.03 & -59469.32 \\
Market + RMW & Fernández & 64 & 29046.23 & -57964.46 & -57397.75 \\
Market + RMW & TabPFN 3.5 & 64 & 29285.19 & -58442.39 & -57875.68 \\
Market + CMA & Fernández & 64 & 29358.31 & -58588.62 & -58021.91 \\
Market + CMA & TabPFN 3.5 & 64 & 29488.86 & -58849.71 & -58283.01 \\
Market + Momentum & Fernández & 64 & 28898.26 & -57668.52 & -57101.81 \\
Market + Momentum & TabPFN 3.5 & 64 & 29152.82 & -58177.65 & -57610.94 \\
FF5 & Fernández & 67 & 32006.28 & -63878.57 & -63285.30 \\
FF5 & TabPFN 3.5 & 67 & 32067.11 & -64000.21 & -63406.94 \\
FF5 + Momentum & Fernández & 68 & 32192.00 & -64248.00 & -63645.87 \\
FF5 + Momentum & TabPFN 3.5 & 68 & 32258.19 & -64380.37 & -63778.25 \\
\bottomrule
\end{tabular}

    \caption{Spatial lag models on the common issuer-quarter mask. Information criteria
    condition on estimated first-stage exposures.}
\end{table*}
  
  \begin{table*}[p]
    \centering
    \fftabletype
    \begin{tabular}{llrrrr}
\toprule
Factors & Peer graph & Parameters & Log likelihood & AIC & BIC \\
\midrule
Market & Fernández & 63 & 28125.66 & -56125.32 & -55567.47 \\
Market & TabPFN 3.5 & 63 & 28374.49 & -56622.98 & -56065.13 \\
SMB & Fernández & 63 & 27734.47 & -55342.93 & -54785.08 \\
SMB & TabPFN 3.5 & 63 & 27972.94 & -55819.88 & -55262.03 \\
HML & Fernández & 63 & 27897.88 & -55669.75 & -55111.90 \\
HML & TabPFN 3.5 & 63 & 27978.53 & -55831.06 & -55273.21 \\
RMW & Fernández & 63 & 27084.24 & -54042.47 & -53484.62 \\
RMW & TabPFN 3.5 & 63 & 27337.03 & -54548.06 & -53990.21 \\
CMA & Fernández & 63 & 27513.15 & -54900.30 & -54342.45 \\
CMA & TabPFN 3.5 & 63 & 27651.44 & -55176.89 & -54619.04 \\
Momentum & Fernández & 63 & 27262.17 & -54398.34 & -53840.48 \\
Momentum & TabPFN 3.5 & 63 & 27536.71 & -54947.42 & -54389.57 \\
Market + SMB & Fernández & 64 & 29173.88 & -58219.76 & -57653.05 \\
Market + SMB & TabPFN 3.5 & 64 & 29394.30 & -58660.59 & -58093.89 \\
Market + HML & Fernández & 64 & 30018.39 & -59908.78 & -59342.07 \\
Market + HML & TabPFN 3.5 & 64 & 30066.34 & -60004.69 & -59437.98 \\
Market + RMW & Fernández & 64 & 29055.23 & -57982.45 & -57415.74 \\
Market + RMW & TabPFN 3.5 & 64 & 29280.94 & -58433.89 & -57867.18 \\
Market + CMA & Fernández & 64 & 29365.54 & -58603.09 & -58036.38 \\
Market + CMA & TabPFN 3.5 & 64 & 29472.72 & -58817.44 & -58250.73 \\
Market + Momentum & Fernández & 64 & 28890.93 & -57653.85 & -57087.15 \\
Market + Momentum & TabPFN 3.5 & 64 & 29139.49 & -58150.98 & -57584.27 \\
FF5 & Fernández & 67 & 32000.96 & -63867.92 & -63274.65 \\
FF5 & TabPFN 3.5 & 67 & 32036.23 & -63938.46 & -63345.19 \\
FF5 + Momentum & Fernández & 68 & 32186.12 & -64236.24 & -63634.11 \\
FF5 + Momentum & TabPFN 3.5 & 68 & 32216.53 & -64297.06 & -63694.93 \\
\bottomrule
\end{tabular}

    \caption{Spatial error models on the common issuer-quarter mask. Information criteria
    condition on estimated first-stage exposures.}
\end{table*}
}{\ffmissing}

\end{document}